\documentclass[aps,prl,reprint,superscriptaddress,longbibliography]{revtex4-2}
\usepackage{amsmath,amssymb}
\usepackage{graphicx}
\usepackage{bm}
\usepackage[colorlinks=true,linkcolor=blue,citecolor=blue,urlcolor=blue]{hyperref}
\newcommand{\De}{\mathrm{De}}
\newcommand{\Kn}{\mathrm{Kn}}
\newcommand{\Ma}{\mathit{M}}
\newcommand{\Rey}{\mathrm{Re}}
\newcommand{\Web}{\mathrm{We}}
\newcommand{\Tpar}{T_{\parallel}}
\newcommand{\Tperp}{T_{\perp}}
\newcommand{\tth}{\tau_\theta}
\begin{document}

\title{Why gas-focused microjets are so fast:\\
kinetically resolved, shear-driven flow focusing in vacuum}
\author{A. M. Ga\~n\'an-Calvo}
\affiliation{Depto.\ de Ingenier\'ia Aeroespacial y Mec\'anica de Fluidos,
Universidad de Sevilla, E-41092 Sevilla, Spain}
\email{amgc@us.es}
\date{\today}

\begin{abstract}
Gas-focused liquid microjets---the flow-focusing sample delivery on which serial femtosecond
crystallography depends---run several times faster than the pressure-driven (Bernoulli) bound, dragged by
the tangential stress of the focusing gas. The magnitude of that stress, and hence the speed, is fixed by
the gas viscosity in a rarefied, hypersonic expansion that continuum, local-equilibrium models assume
rather than resolve. We resolve it with a deterministic kinetic (Shakhov--BGK) solver coupled to the
slender liquid jet. The jet is \emph{shear-driven}, not pressure-driven: a first integral of the momentum
balance shows the tangential stress supplying nearly all of the axial momentum, the pressure term being
bounded by the Bernoulli value. The gas does
not become ballistic behind the near field---its stress decays as a power law and it stays coupled---and its
constitutive regime is set by a single rarefaction parameter $\delta=D/\ell_0$, the orifice diameter over the
source mean free path, through the thermodynamic Deborah number $\De_\theta\simeq\Kn\,\Ma$ (Knudsen times
Mach), whose $\De_\theta=1$ surface maps where the Newtonian-gas closure fails: the small-$\delta$ vacuum
corner where crystallography jets operate. The kinetically computed surface stress is the input for the
fully non-Newtonian (viscoelastic-liquid) sequel.
\end{abstract}
\maketitle

\emph{Introduction.}---Serial femtosecond crystallography (SFX) reconstructs macromolecular structure from
millions of diffraction snapshots of a hydrated sample stream intersected by femtosecond X-ray pulses
\cite{Chapman2011}. The stream is a gas-focused liquid microjet: the \emph{flow focusing} principle
introduced by Ga\~n\'an-Calvo \cite{GananCalvo1998}, in which a co-flowing gas discharged through a small
orifice compresses a liquid meniscus into a micron jet by its aerodynamic stress alone, without a solid
nozzle wall \cite{GananCalvo1998,Rubio2021,Montanero2024}. Its adaptation to sample injection into
vacuum---the ``gas dynamic virtual nozzle'' (GDVN) \cite{DePonte2008,Weierstall2012,Knoska2020}---rests on
this same mechanism, which its originators explicitly attribute to Ref.~\cite{GananCalvo1998} in the
opening statement of Ref.~\cite{DePonte2008}. The serial interrogation of discrete units---cells, viruses,
and \emph{microcrystals}---carried one by one in such a gas-focused capillary jet, with an unbroken jet
many diameters long, was patented for this purpose in 2003 \cite{GananCalvoPatent2003}, anticipating the
sample-delivery concept SFX would later require. Two requirements are severe and simultaneous: the jet
must be thin, of order a micron, to fit the focus, and it must be \emph{fast}, so that at
megahertz-repetition facilities it clears the interaction volume between pulses.

\begin{figure}[h]
\includegraphics[width=0.8\columnwidth]{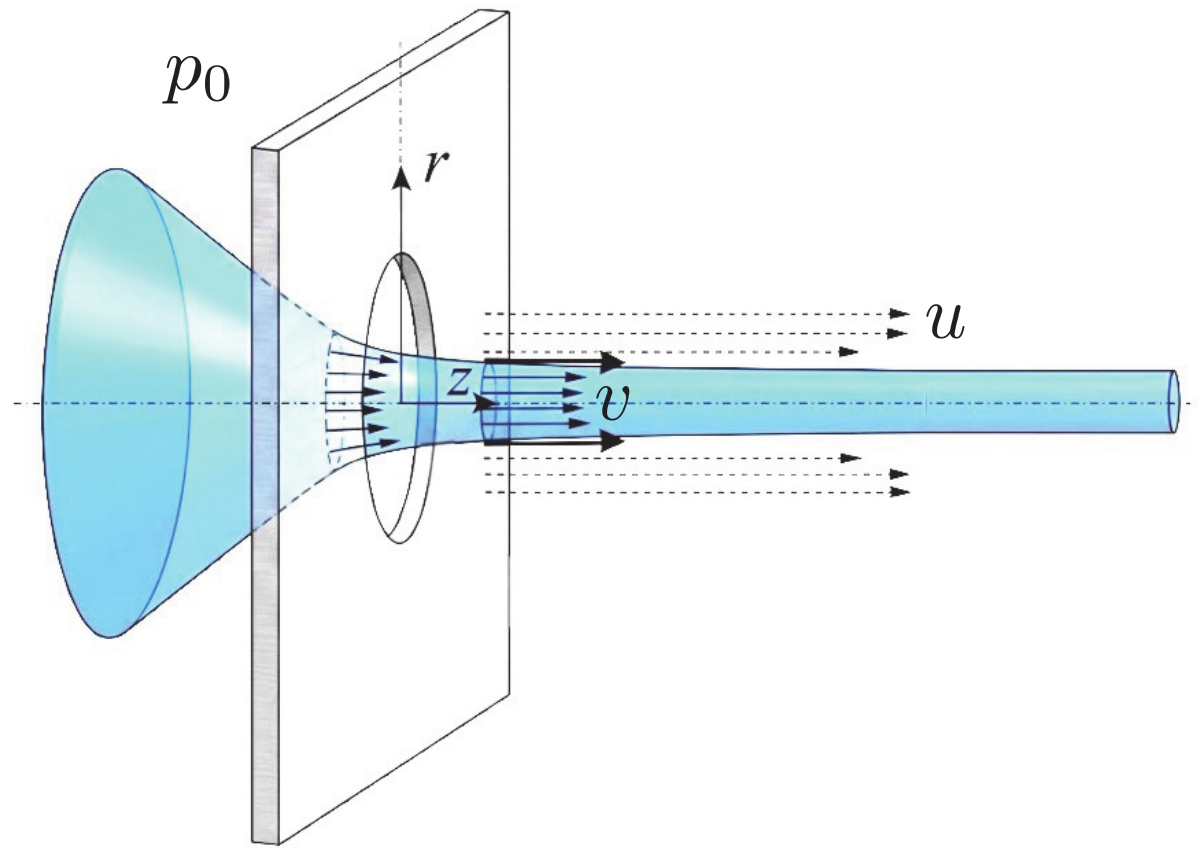}
\caption{Sketch of gas-focused flow focusing: the liquid (blue) issues from the stagnation chamber (pressure $p_0$) through a thin-plate orifice of diameter $D$, and the co-flowing gas, expanding into vacuum with velocity $\bm u$ (dashed arrows), drives the liquid jet of velocity $v$ (solid arrows). Axial and radial coordinates $(z,r)$.}
\label{fig:FF1}
\end{figure}

These speeds exceed what the gas pressure drop alone could deliver. A Bernoulli balance
$\tfrac12\rho_\ell v_{\max}^2=\Delta p=p_0-p_b$---with $p_b$ the ambient back pressure into which the
orifice discharges---caps the pressure-driven liquid at $v_{\max}=U^\ast\sqrt{1-p_b/p_0}$, with the
gas-momentum velocity $U^\ast\equiv\sqrt{2p_0/\rho_\ell}=v_m\sqrt{\rho_0/\rho_\ell}$
($v_m=\sqrt{2k_BT_0/m}$ the most probable molecular speed, $k_B$ the Boltzmann constant, $m$ the molecular
mass, $\rho_0$ the stagnation gas density); in vacuum ($p_b\to0$) that bound is exactly $v/U^\ast=1$, and
gas-focused jets run several times faster. That they do is not, in itself, a puzzle, and has long been
understood in qualitative terms: the liquid is not pushed by its own pressure drop but dragged by a gas
streaming past it at speeds of order $v_m$, and it is the tangential viscous stress of that gas that
transfers the axial momentum \cite{DePonte2008,GananCalvoFerrera2011,Rubio2021}. What has never been
established is the \emph{magnitude} of that stress---and the magnitude is what fixes the speed. It is set
by the gas viscosity at the strongly expanded state, through the interfacial shear rate; determining it is
therefore a constitutive question about a gas that reaches speeds of order $v_m$ and loses orders of
magnitude of density within a few orifice diameters. Continuum models of the focusing gas integrate the
two-phase compressible Navier--Stokes equations with a Newtonian, local-equilibrium (LTE) closure
\cite{Zahoor2018,Rubio2021,HerradaMontanero2016}; they can match jet diameters and lengths, but they are
not thermodynamically exact for the gas: the LTE closure misassigns the gas viscosity at the strongly
expanded state, and as $\delta\to1$ or below the gas freezes and its effective viscosity departs from the
equilibrium value that such models assign. They treat the gas as being in local equilibrium everywhere. That
assumption has lately been examined for this very system, and found safe: Zahoor \emph{et al.}
\cite{Zahoor2025} obtain $\Kn=0.0044$ from the orifice opening and the reference gas density, while
Kova\v{c}i\v{c} \emph{et al.} \cite{Kovacic2024} map the Knudsen number of the helium expansion into
vacuum and conclude that the Navier--Stokes treatment ``is justified''. Both conclusions hold within their
stated scope---the first model is incompressible and isothermal with an atmospheric outlet; the second
assigns the expanded plume the vacuum-chamber diameter, $L\simeq300$~mm, as its length scale. But the
criterion applied, $\Kn<0.01$, is not the one that governs a hypersonic expansion: local equilibrium fails
when the strain rate outruns the collision rate, $\De_\theta\simeq\Kn\,\Ma\to1$ \cite{Tsien1946}, and it is
the Mach factor that is decisive. The values reported in Ref.~\cite{Kovacic2024} itself ($\Ma\simeq6$, and
$\Kn$ up to $0.116$ at the orifice lip) already give $\De_\theta\simeq0.7$---at the freezing threshold---in
cells that a $\Kn$-only test discards. Downstream, where the stress is actually delivered, both the mean
free path and $\Ma$ grow and $\De_\theta$ rises steeply, the mechanism Ref.~\cite{Kovacic2024} itself
anticipates in words: the gas expansion causes ``the mean free path of helium molecules to increase, where
rarefied gas effects could become non-negligible''. Where the gas rarefies, the LTE closure fails
\cite{Tsien1946,Bird1970,Boyd1995,MohanGirimaji2022}, and it is precisely the deviatoric (shear) stress of
that gas that is applied, tangentially, to the liquid surface (Figure \ref{fig:FF1}). We therefore resolve the focusing gas in
the kinetic regime, with a deterministic solver of the Boltzmann model equation, and couple the resulting
surface stress to the liquid jet.

\emph{Kinetic gas model.}---We solve the steady, axisymmetric kinetic model equation of Shakhov
\cite{Shakhov1968} ($\mathrm{Pr}=2/3$ the Prandtl number; the pure Bhatnagar--Gross--Krook (BGK) limit
is $\mathrm{Pr}=1$) for a monatomic gas
($\gamma=5/3$ the adiabatic index, zero bulk viscosity) expanding from a stagnation reservoir $(p_0,T_0)$ through a thin-plate
orifice of diameter $D$ into vacuum. The gas viscosity follows the power law
$\mu\propto T^{\omega}$ ($\omega=0.66$ for helium), which sets the temperature dependence of the collision
rate in Eq.~\eqref{eq:kinetic}. The unknown is the single-particle velocity distribution function
$f(\bm r,\bm c)$, a number density in phase space whose velocity moments are the macroscopic gas fields.
Resolving $f$ itself is what makes the computation closure-free: continuum models do not solve for $f$ but
\emph{assume} it stays close to a local Maxwellian---the assumption that fails as the gas rarefies. We reserve $\bm c$ for the
\emph{molecular} velocity (the argument of $f$) and $\bm u$ for the \emph{macroscopic} gas velocity (a
moment of $f$), keeping $v$ for the liquid. In variables scaled by $D$, $v_m$ and $n_0=p_0/k_BT_0$
($\xi=z/D$, $\hat r=r/D$, $\hat{\bm c}=\bm c/v_m$, $\hat n=n/n_0$, $\hat T=T/T_0$,
$\hat f=f\,v_m^3/n_0$),
\begin{equation}
\hat c_z\,\partial_\xi \hat f+\frac{1}{\hat r}\partial_{\hat r}(\hat r\,\hat c_r\,\hat f)
-\frac{1}{\hat r}\partial_\alpha(\hat c_\theta\,\hat f)=\delta\,\hat n\,\hat T^{1-\omega}\,(\hat f^{S}-\hat f),
\label{eq:kinetic}
\end{equation}
where $\hat c_z$, $\hat c_r$ and $\hat c_\theta$ are the axial, radial and azimuthal components of the
molecular velocity $\hat{\bm c}$; $\hat c_z$ is the one that transports $\hat f$ along the jet axis, while
the other two are written in polar form, $(\hat c_r,\hat c_\theta)=\hat c_{\mathcal R}(\cos\alpha,\sin\alpha)$,
with $\hat c_{\mathcal R}$ their polar magnitude and $\alpha$ their polar angle in the
$(\hat c_r,\hat c_\theta)$ plane, so that $\hat{\bm c}=(\hat c_z,\hat c_{\mathcal R},\alpha)$ in the
representation used by the solver. The collision term relaxes $\hat f$ toward the Shakhov
reference distribution
$\hat f^{S}=\hat f_M\big[1+\tfrac{4}{5}(1-\mathrm{Pr})(\hat{\bm c}-\hat{\bm u})\!\cdot\!\hat{\bm q}\,
(|\hat{\bm c}-\hat{\bm u}|^2/\hat T-\tfrac52)/(\hat n\hat T^{2})\big]$, with
$\hat f_M(\hat n,\hat{\bm u},\hat T)=\hat n\,\pi^{-3/2}\hat T^{-3/2}\,e^{-|\hat{\bm c}-\hat{\bm u}|^2/\hat T}$
the local Maxwellian and $\hat{\bm q}$ the heat-flux vector (both moments of $\hat f$); $\mathrm{Pr}=2/3$
sets the correct thermal conductivity and $\hat f^{S}\!\to\!\hat f_M$ recovers BGK at $\mathrm{Pr}=1$. The
\emph{single} control parameter is the rarefaction number
\begin{equation}
\delta=\frac{D}{\ell_0}=\frac{1}{\Kn_0},
\label{eq:delta}
\end{equation}
the inverse source Knudsen number ($\ell_0$ the stagnation mean free path). Equation~\eqref{eq:kinetic} is
discretised in discrete ordinates on a physical $(\xi,\hat r)$ mesh, the molecular velocity being
represented in polar coordinates $(\hat c_z,\hat c_{\mathcal R},\alpha)$. The angular
(centrifugal/Coriolis) transport in $\alpha$---a velocity-space, not a physical, coordinate---is cast as a
conservative flux, with the cell-averaged radial velocity defined as the difference of the azimuthal-flux
edge values, $\bar c_{r,k}=(\hat c_{\theta,k+1/2}-\hat c_{\theta,k-1/2})/\Delta\alpha$. With this choice the
geometric terms cancel \emph{identically} for $\hat f=\mathrm{const}$, so a uniform gas is preserved to
machine precision; it is the axisymmetric counterpart of Carlson's angular-differencing recursion in
curvilinear-geometry discrete-ordinates ($S_N$) neutron transport, which enforces exactly the same
consistency. Source iteration with positive upwind sweeps converges the system, with continuation in
$\delta$ at large optical thickness. The scheme reproduces the exact collisionless axis solution, the
reduced mass flow---the orifice mass flow rate normalised by its free-molecular (effusive) value,
$W\equiv\dot m/\dot m_{\rm fm}$ with $\dot m_{\rm fm}=p_0A/\sqrt{2\pi R_gT_0}$---tending to $W=1$ as it
must; and it returns a uniform Maxwellian to a residual that is pure velocity-quadrature error under
refinement. Its moments give the number density $\hat n$, the macroscopic gas
velocity $\hat{\bm u}$ and the temperature $\hat T$, the pressure tensor $P_{ij}$ (hence $\Tpar,\Tperp$),
and the thermodynamic Deborah number
\begin{equation}
\De_\theta=\tth\,\dot\varepsilon=\frac{\mu}{p}\,\frac{u}{D}\simeq\Kn\,\Ma ,
\label{eq:Deth}
\end{equation}
with $\tth$ the thermodynamic relaxation time (of the order of the collision time), $\dot\varepsilon$ the
strain rate of the expansion, $p$ the local gas pressure, $u=|\bm u|$ the local gas speed, $\Kn$ the local
Knudsen number and $\Ma=u/a$ the local Mach number ($a=\sqrt{\gamma k_BT/m}$ the sound speed). $\De_\theta$
is thus a Deborah number in the strict sense: the expansion strains the gas faster than collisions can
relax it whenever $\De_\theta\gtrsim1$. This is
Tsien's rarefaction parameter \cite{Tsien1946}: LTE for $\De_\theta\ll1$, onset near $\De_\theta\simeq0.05$
\cite{Bird1970,Boyd1995}, frozen for $\De_\theta\gtrsim1$. On a free-jet axis $\Ma\propto(z/D)^{\gamma-1}$
and $n\propto(z/D)^{-2}$ \cite{AshkenasSherman1966}, so $\De_\theta(z)\propto\delta^{-1}(z/D)^{\gamma+1}$
(exponent $8/3$ for $\gamma=5/3$): small at the orifice, but rising steeply downstream and inversely with
$\delta$.

\emph{Coupling to the slender jet.}---From the kinetic solution we extract the two gas fields that act on
the liquid: the gas pressure $p_g(z)$ and the tangential (shear) stress $\tau_s(z)$ exerted on the jet
surface. Both are normalised by the stagnation pressure, $\hat p_g=p_g/p_0$ and $\hat\tau_s=\tau_s/p_0$,
and are obtained by solving Eq.~\eqref{eq:kinetic} around a coaxial thread (radius $\ll D/2$; in that thin-thread
limit the fields are insensitive to the thread radius). The stress $\hat\tau_s$ is the wall shear the
kinetic gas exerts on the interface, a diffusely reflecting (fully accommodating, $\sigma=1$) surface
moving at the local jet velocity: the
re-emitted molecules carry the wall momentum, so the stress is set by the gas velocity \emph{relative} to
the jet. This is retained exactly through $\hat\tau_s(z)=\hat\tau_s^{\rm rest}(z)-2\beta\,\hat\Phi(z)\,\hat v$,
with $\hat\tau_s^{\rm rest}$ the stress on a stationary surface, $\hat\Phi$ the incident mass flux (both
from the same kinetic solution), and $\beta=\sqrt{\rho_0/\rho_\ell}=U^\ast/v_m$ the (small) gas-to-liquid
density-ratio parameter; retaining the jet motion lowers the terminal speed by only a few percent. Because
the kinetic solver is practicable at rarefaction below the operating $\delta$, we verified that the coupled
result is robust to $\delta$: the normalised stress $\hat\tau_s^{\rm rest}(z/D)$ is nearly self-similar over
the accessible range, and the shear dominance established below not only persists but \emph{strengthens}
toward the continuum, where the pressure share falls. Grid convergence, $\delta$-sensitivity, the inlet
attractor and the moving-wall correction are quantified in the Supplemental Material. The slender jet of
radius $R(z)$ and speed $v(z)=Q/(\pi R^2)$, fed at the constant volumetric flow rate $Q$, obeys the quasi-one-dimensional momentum balance
\cite{GananCalvoFerrera2011} with the exact elongational viscous term of the one-dimensional Navier--Stokes
reduction \cite{Eggers1993,EggersDupont1994}. Scaling $R$ by $D/2$ ($y=2R/D$), $v$ by
$U^\ast=v_m\sqrt{\rho_0/\rho_\ell}$ ($\hat v=v/U^\ast=\Lambda/y^2$, $\Lambda=4Q/\pi D^2U^\ast$ the
dimensionless feed rate) and $z$ by $D$, it becomes
\begin{equation}
\hat v\,\frac{d\hat v}{d\xi}=-\tfrac12\,\frac{d\hat p_g}{d\xi}
-\frac{2}{\Web}\frac{d}{d\xi}\!\Big(\frac1y\Big)
+\frac{2\hat\tau_s}{y}
+\frac{3}{\Rey}\frac{1}{y^2}\frac{d}{d\xi}\!\Big(y^2\frac{d\hat v}{d\xi}\Big),
\label{eq:jet}
\end{equation}
governed by a liquid Reynolds and Weber number built on $U^\ast$ and $D$,
\begin{equation}
\Rey=\frac{\rho_\ell U^\ast D}{\mu_\ell},\qquad \Web=\frac{\rho_\ell U^{\ast2}D}{\gamma_s},
\label{eq:ReWe}
\end{equation}
with $\mu_\ell$ the liquid viscosity and $\gamma_s$ its surface tension (subscripted to avoid collision
with the adiabatic index $\gamma$); the pressure coefficient $p_0/\rho_\ell U^{\ast2}=\tfrac12$ is fixed by
the choice of $U^\ast$. The five
terms are inertia, gas pressure gradient, surface tension, gas tangential stress, and the Trouton viscous
stress; the inviscid limit $\Rey\to\infty$ collapses the last term. Equation~\eqref{eq:jet} is a
second-order boundary-value problem (jet fed at $y=1$ upstream, $d\hat v/d\xi\to0$ downstream); only when
the pressure term acts alone does it integrate to Bernoulli, $\hat v\to1$.

\begin{figure}[t]
\includegraphics[width=\columnwidth]{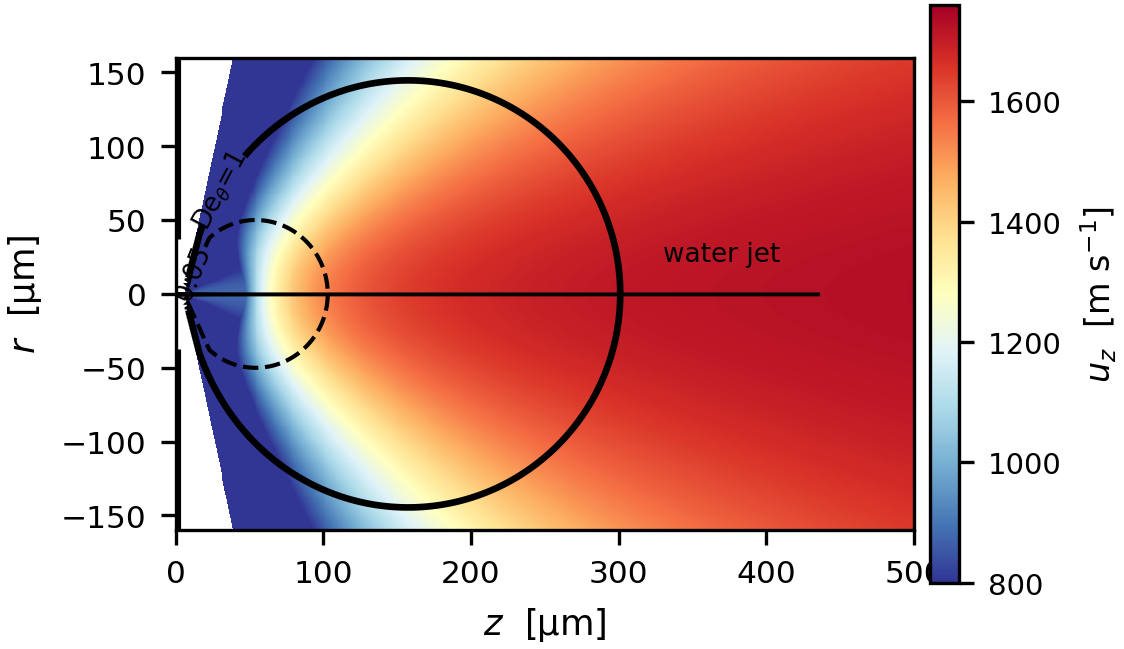}
\caption{Illustrative operating point (helium, $D=75~\mu$m, $p_0\simeq1.2$~bar, $T_0=22^\circ$C; water at
$Q=20~\mu$L/min, i.e.\ $\Rey\simeq10^{3}$, $\Web\simeq2.6\times10^{2}$, $\delta\simeq5\times10^{2}$). Axial gas
velocity $u_z$ (free-jet field at this operating point, where the near field is continuum) with the liquid jet (black, on axis) and the Deborah contours $\De_\theta=0.05$
(dashed) and $\De_\theta=1$ (solid, sudden-freeze surface). The jet axis crosses the solid contour at the
freezing distance $(z/D)_f$; inside it the gas is quasi-Newtonian, outside frozen and non-Newtonian. This
single realization is illustrative; the physics is governed by the dimensionless groups
$(\Rey,\Web;\delta)$ swept in Figs.~\ref{fig:freeze} and \ref{fig:jet}.}
\label{fig:map}
\end{figure}

\begin{figure}[t]
\includegraphics[width=0.9\columnwidth]{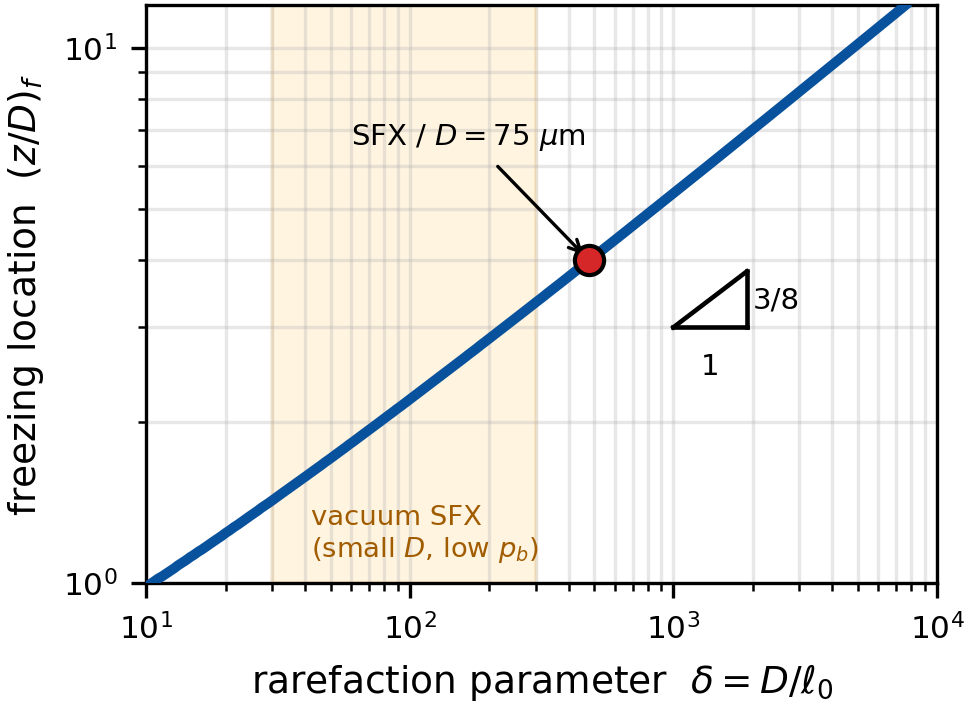}
\caption{Freezing distance $(z/D)_f$ (where $\De_\theta=1$ on the axis) versus the rarefaction parameter
$\delta=D/\ell_0$, following $\delta^{3/8}$ (dotted). Small $\delta$ (small $D$, low back pressure) is the
vacuum-SFX corner, where the gas freezes within a diameter or two; the marker is the representative
operating point of Fig.~\ref{fig:map}.}
\label{fig:freeze}
\end{figure}

\begin{figure*}[t]
\includegraphics[width=\textwidth]{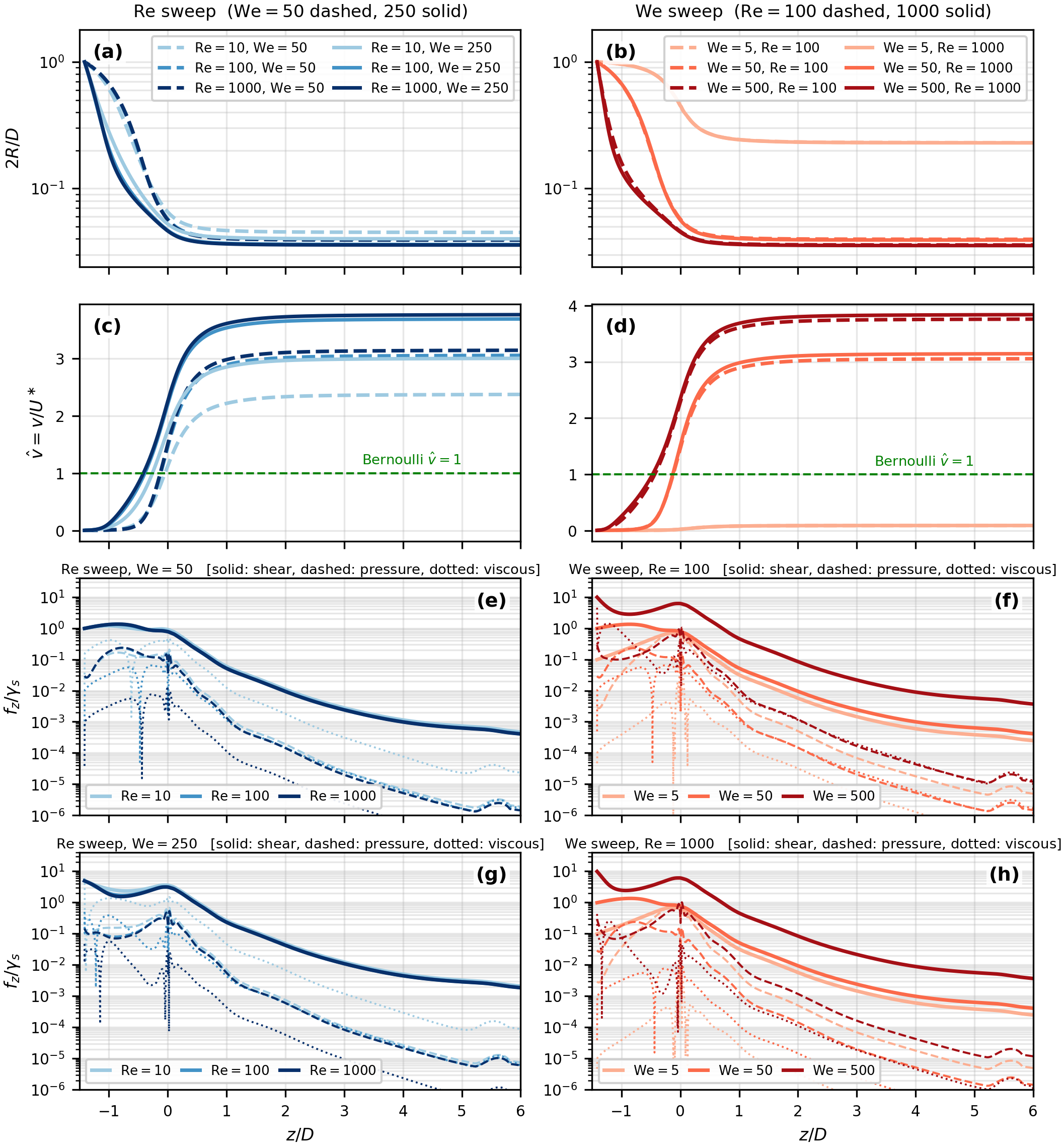}
\caption{Dimensionless jet over $(\Rey,\Web)$ (gas fixed, $\delta=1$). Left column, $\Rey$ sweep;
right column, $\Web$ sweep; each curve family is shown at two values of the other group, so the response is
sampled over the whole plane. (a,b) jet diameter $2R/D$ (log) and (c,d) speed $\hat v=v/U^\ast$, with six
curves each: colour denotes the swept group, line style the fixed one [(a,c) $\Web=50$ dashed, $250$ solid;
(b,d) $\Rey=100$ dashed, $1000$ solid]. The pressure-only Bernoulli bound $\hat v=1$ (green) is exceeded
several-fold in every case except the capillary-arrested $\Web=5$. (e--h) axial force per unit length scaled
by surface tension $\gamma_s$, split by fixed group to keep the traces legible (solid: gas shear; dashed: gas pressure;
dotted: Eggers viscous): (e) $\Rey$ sweep at $\Web=50$, (g) at $\Web=250$; (f) $\Web$ sweep at $\Rey=100$,
(h) at $\Rey=1000$. The gas shear dominates the other two by one to two decades throughout the focusing
region, for \emph{every} $(\Rey,\Web)$: the jet is shear-driven across the whole parameter plane. (The
downward spikes in the dotted traces are sign changes of the viscous term on a logarithmic axis.)}
\label{fig:jet}
\end{figure*}

\emph{General behaviour.}---The parameter space is $(\Rey,\Web;\delta)$. Figure~\ref{fig:map} shows the
kinetic axial velocity field with the jet and the $\De_\theta=0.05$ and $\De_\theta=1$ contours: the latter
closes around the near-orifice region, the jet axis crossing it at the freezing distance $(z/D)_f$. By
Eq.~\eqref{eq:Deth}, $\De_\theta=1$ on the axis at $(z/D)_f\propto\delta^{1/(\gamma+1)}=\delta^{3/8}$ ($\gamma=5/3$); Fig.~\ref{fig:freeze} maps it. Vacuum-SFX operation (small $D$, low back pressure)
occupies small $\delta$, where the gas freezes within a diameter or two and the standard Newtonian-gas
closure is inapplicable; laboratory atmospheric-discharge focusing occupies large $\delta$, where it is
adequate. The liquid response over $(\Rey,\Web)$ is shown in Fig.~\ref{fig:jet}: the jet focuses from the
meniscus to a terminal $2R/D$ that decreases, and a terminal $\hat v$ that increases, with both $\Rey$
(less viscous drag) and $\Web$ (weaker capillary resistance); at low $\Web$ the capillary force arrests the
focusing altogether.

\emph{Result 1: the jet is shear-driven.}---The statement is exact, and independent of the operating point.
A first integral of Eq.~\eqref{eq:jet} gives the terminal energy budget
\begin{equation}
\tfrac12\hat v_\infty^{2}-\tfrac12\hat v_0^{2}
=\underbrace{\tfrac12\Delta\hat p_g}_{\le\,1/2}
+\;\mathcal S\;
-\frac{2}{\Web}\Big[\frac1y\Big]_0^{\infty}+\mathcal V,
\,
\mathcal S\equiv\int_0^{\infty}\frac{2\hat\tau_s}{y}\,d\xi ,
\label{eq:budget}
\end{equation}
with $\mathcal V$ the Trouton viscous term. Since $\Delta\hat p_g\le1$ in vacuum, the pressure term can
never exceed $\tfrac12$: \emph{the Bernoulli bound $\hat v=1$ is precisely the $\mathcal S\to0$ limit of
Eq.~\eqref{eq:budget}}, and the whole of the excess over that bound is carried by the dimensionless shear
functional $\mathcal S$, with $\hat v_\infty\simeq(1+2\mathcal S)^{1/2}$. That $\mathcal S\gg1$ is
geometric, and equally free of the operating point: the ratio of the two gas terms in Eq.~\eqref{eq:jet} is
\begin{equation}
\frac{F_{\rm shear}}{F_{\rm press}}=\frac{2\pi R\,\tau_s}{\pi R^{2}|dp_g/dz|}
=\frac{4\hat\tau_s}{y\,|d\hat p_g/d\xi|}=O(y^{-1}),
\label{eq:ratio}
\end{equation}
because the pressure gradient acts on the vanishing cross-section $\pi R^{2}$ while the shear acts on the
perimeter $2\pi R$: shear is amplified over pressure by $D/2R=1/y\gg1$ for \emph{any} thin jet.
Fig.~\ref{fig:jet}(e,f) bears this out---the shear exceeds both the pressure and the Eggers viscous term by
one to two decades across the focusing region, for every $(\Rey,\Web)$---and the inviscid and viscous jets
of Fig.~\ref{fig:jet}(a) differ imperceptibly. For the illustrative point of Fig.~\ref{fig:map} the four
terms of Eq.~\eqref{eq:budget} are $0.49$ (pressure, i.e.\ at the Bernoulli ceiling), $\mathcal S=6.9$
(shear), $-0.22$ (capillary) and $-0.02$ (viscous), giving $\hat v_\infty\simeq3.8$: several times the
pressure-only bound (Fig.~\ref{fig:jet}(c,d)). This is what sets the speed: the
extraordinary speeds of gas-focused SFX jets are the accumulated tangential impulse of a gas moving at
$\sim v_m$ past the surface, not pressure work.

\begin{figure}[h]
\includegraphics[width=0.9\columnwidth]{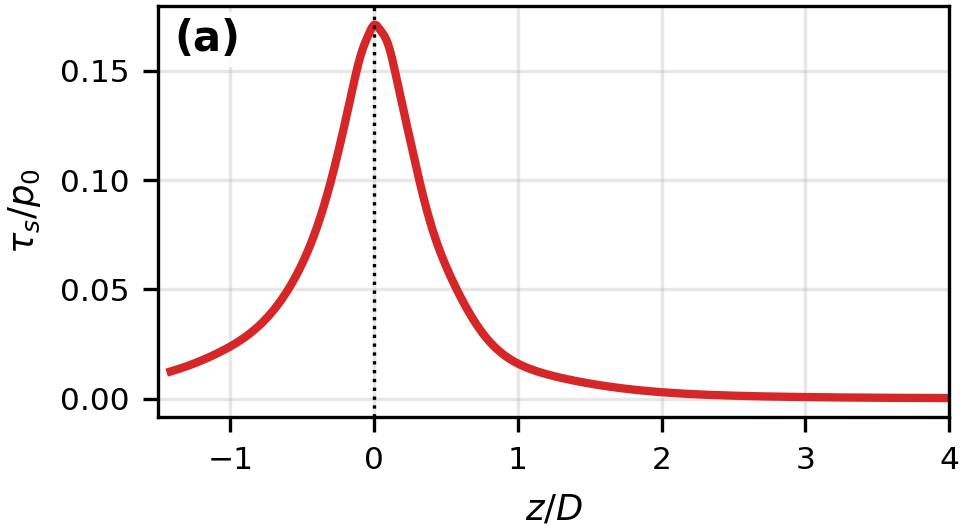}\\
\includegraphics[width=0.9\columnwidth]{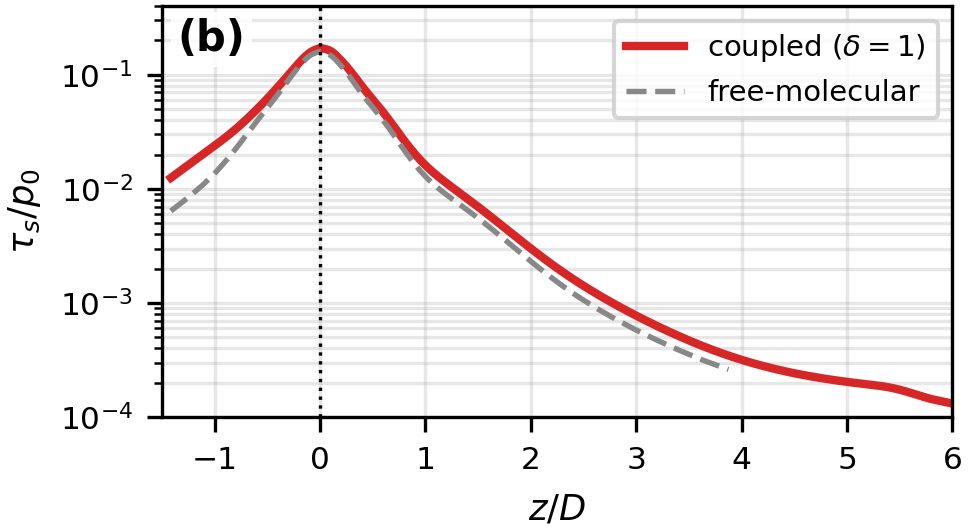}\\
\includegraphics[width=0.9\columnwidth]{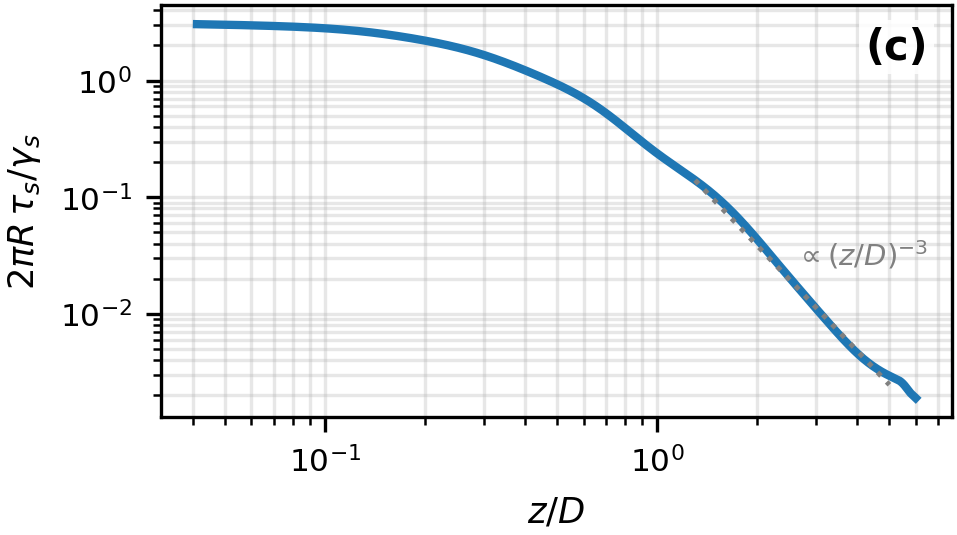}\\
\includegraphics[width=0.9\columnwidth]{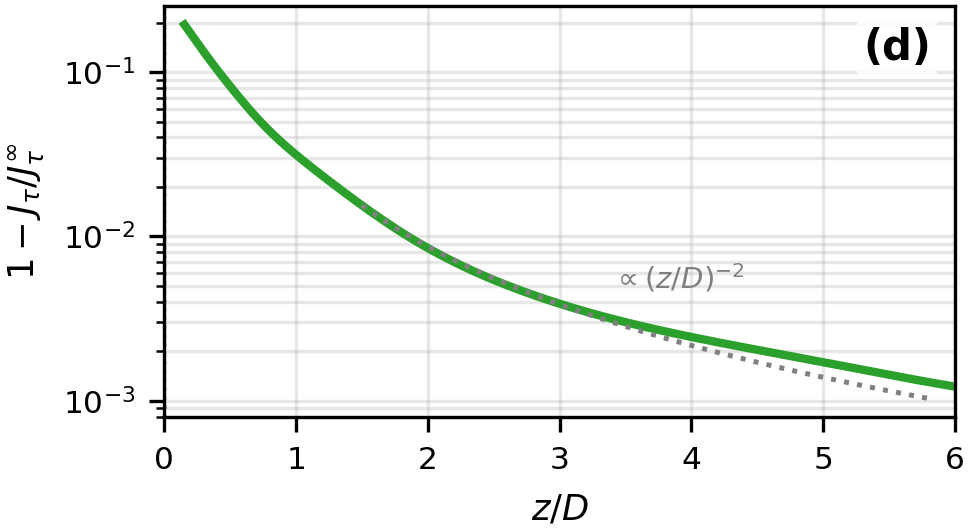}
\caption{Gas tangential stress on the jet, for the illustrative operating point of Fig.~\ref{fig:map}
($\Rey\simeq10^{3}$, $\Web\simeq2.6\times10^{2}$; gas at the $\delta$ accessible to the kinetic solver).
(a) $\tau_s/p_0$; (b) its logarithmic decay, coupled versus free-molecular, both a power-law tail rather
than a cut-off; (c) traction per unit length $2\pi R\tau_s/\gamma_s$ (log--log), following
$\propto(z/D)^{-3}$; (d) impulse deficit $1-J_\tau/J_\tau^\infty$ (log), an asymptotic power law
$\propto(z/D)^{-2}$---the gas keeps delivering momentum downstream.}
\label{fig:shear}
\end{figure}

\emph{Result 2: the gas is coupled, not ballistic.}---Figure~\ref{fig:shear} shows $\tau_s/p_0$ peaking at
the orifice and decaying as a \emph{power law}, not a cut-off; the coupled and free-molecular limits
bracket it. The tangential impulse is concentrated within a diameter or two---because the density falls as
$(z/D)^{-2}$, not because the gas ceases to interact: writing the cumulative tangential impulse delivered to the jet
as $J_\tau(z)=\int_0^{z}2\pi R\,\tau_s\,dz'$ and its asymptotic total as $J_\tau^\infty$, the deficit
$1-J_\tau/J_\tau^\infty$ decays as a power law (Fig.~\ref{fig:shear}(d)), and the gas continues to stream past the jet at a
speed of order $v_m$, an order of magnitude above the liquid. The slip, and hence the shear, never vanish.
Treating the jet as ballistic behind the near field discards this tail and the physical reason the coupling
exists; a rarefied, even frozen, gas is not a decoupled one---it retains its directed momentum and keeps
delivering stress \cite{HamelWillis1966,ToenniesWinkelmann1977}.

\begin{figure}[h]
\includegraphics[width=0.9\columnwidth]{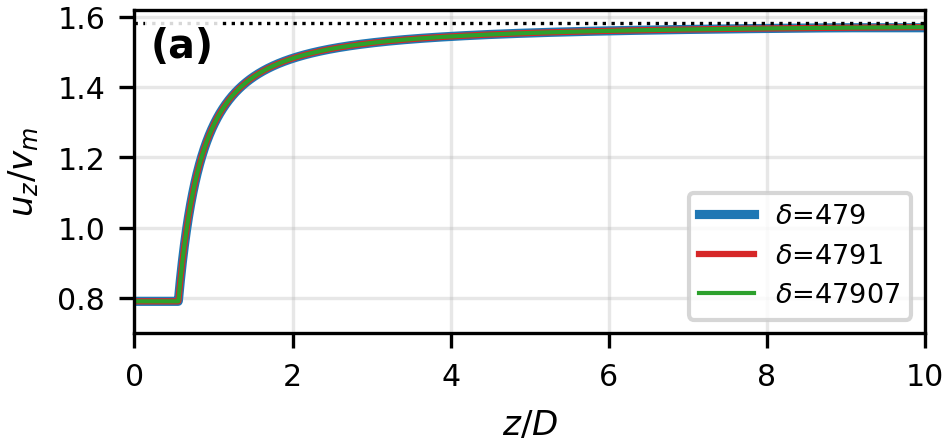}\\
\includegraphics[width=0.9\columnwidth]{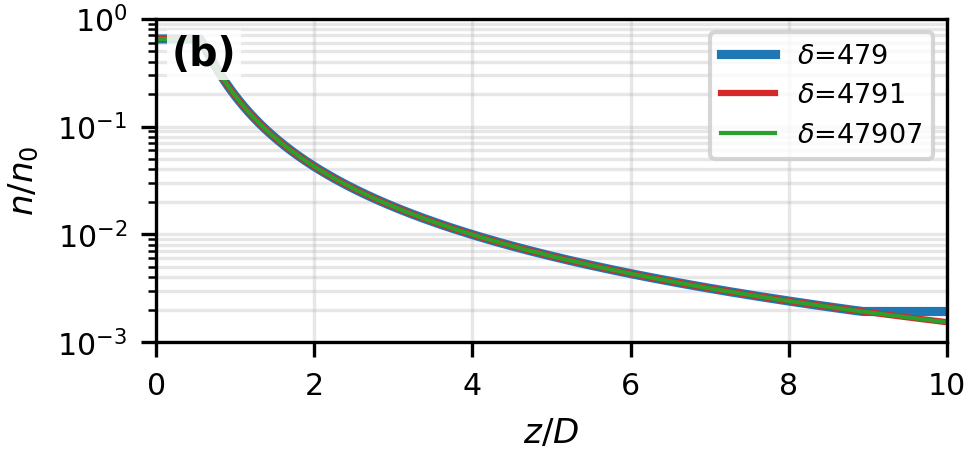}\\
\includegraphics[width=0.9\columnwidth]{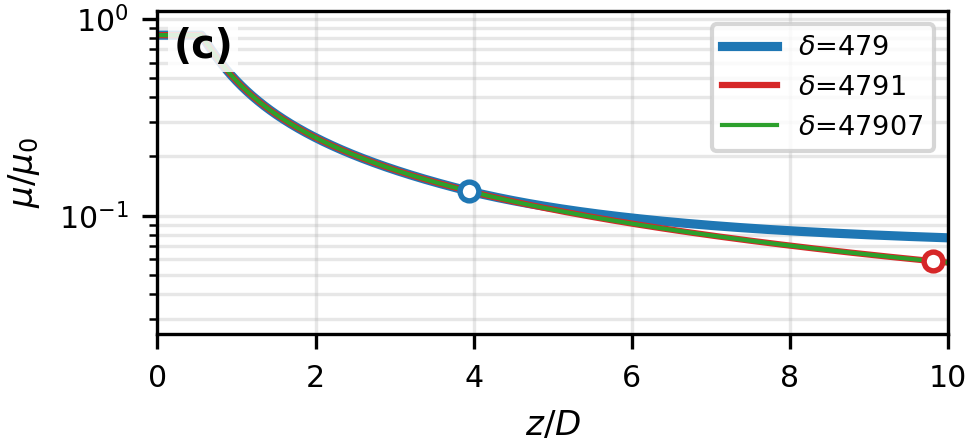}\\
\includegraphics[width=0.9\columnwidth]{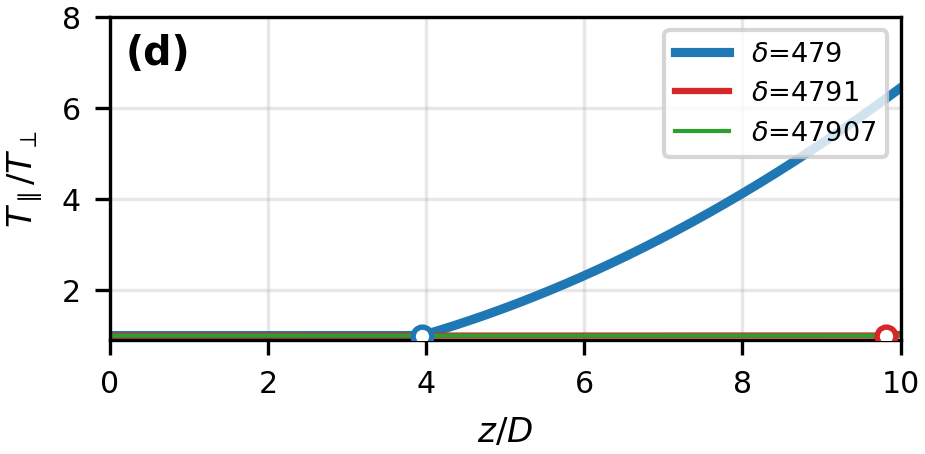}
\caption{Free-jet expansion with the classical two-temperature (sudden-freeze) closure, validated by the kinetic solver in the accessible transitional range. Same stagnation state, three rarefaction parameters $\delta=D/\ell_0$. (a) Axial velocity
$u_z/v_m$ and (b) density $n/n_0$ collapse in $z/D$; (c) gas viscosity $\mu/\mu_0$ and (d) anisotropy
$\Tpar/\Tperp$ do not: they are set by $\De_\theta\propto\delta^{-1}$. Open circles: the $\De_\theta=1$
freezing surface. The Newtonian-gas closure holds at large $\delta$ and fails toward small $\delta$
(vacuum SFX).}
\label{fig:rarefaction}
\end{figure}

\emph{Result 3: rarefaction sets the regime.}---At fixed stagnation state the kinematic fields collapse in
$z/D$ (Fig.~\ref{fig:rarefaction}(a,b)): $u_z/v_m$ and $n/n_0$ are the same for widely different $\delta$.
What does not collapse is the rarefaction: $\De_\theta\propto\delta^{-1}$, so the gas viscosity and the
translational anisotropy $\Tpar/\Tperp$ (Fig.~\ref{fig:rarefaction}(c,d)) are the only fields that
distinguish the orifices, the smallest $\delta$ (most rarefied) departing first. The open circles mark the
sudden-freeze surface $\De_\theta=1$ (the axis crossing of the solid contour of Fig.~\ref{fig:map}), beyond
which the two-temperature closure freezes $\Tpar$; the sharp break is an idealization of the sudden-freeze
switch. The two-temperature anisotropy and its freezing surface are classical rarefied-gas physics
\cite{AndersonFenn1965,HamelWillis1966,ToenniesWinkelmann1977,Miller1988}, mapped directly in free jets by
Raman spectroscopy \cite{Tejeda1996,Montero2013}; what this work adds is not the freezing but its coupling
to the liquid jet---the gas keeps delivering shear as it crosses that surface.

\emph{Relation to prior work.}---The gas-focused microjet has been computed before as a two-phase
compressible Navier--Stokes problem with a Newtonian, local-equilibrium gas
\cite{Zahoor2018,Rubio2021,HerradaMontanero2016}, reproducing jet diameters and lengths; such models even
report the jet ``not affected by the strong temperature and viscosity changes in the focusing gas''
\cite{Zahoor2018}. Those computations contain the interfacial shear implicitly, but do not decompose the
momentum---so the shear-driven mechanism was not identified---and apply the continuum closure throughout,
including where $\De_\theta=O(1)$ and its error is $O(\De_\theta)$, i.e.\ leading order, in the small-$\delta$
vacuum corner where SFX operates. Being Newtonian--LTE, they moreover assign the gas its equilibrium
viscosity $\mu(T)$ at the expanded state, whereas the frozen gas carries a different, anisotropic deviatoric
response; since the surface stress is set by $\mu$ times the shear rate, this misassignment
falsifies the very stress that focuses the jet as $\delta\to O(1)$. The rapid \emph{absolute} decay of $\tau_s$ is not a smallness: the
perimeter-to-area geometry makes the shear the dominant momentum term even as its magnitude falls. The
present computation is closure-free---it resolves the gas kinetically---and $\De_\theta(\delta)$ delimits
the boundary of validity of the standard model, with the surface stress evaluated on the kinetic side of it. It thereby resolves a standing discrepancy that continuum, local-equilibrium models structurally cannot.

\emph{Discussion.}---Three statements, one mechanism: the gas-focused jet is accelerated by the gas
tangential stress, not its pressure; that stress is delivered by a hypersonic gas that stays coupled to the
jet past the near field; and the constitutive regime of that gas is set by $\delta$. Together they account,
from the kinetic equation upward, for jet speeds far above the pressure-driven bound. As a representative
operating point (helium, $D=75~\mu$m, $p_0\simeq1.2$~bar, water at $Q=20~\mu$L/min; $\Rey\simeq10^{3}$,
$\Web\simeq2.6\times10^{2}$, $\delta\simeq5\times10^{2}$) the model gives a micron-scale jet at
$\hat v\simeq3.8$, i.e.\ tens of metres per second, against the $\hat v=1$ pressure bound. The predictions are
falsifiable: the terminal $2R/D$ and $\hat v$ track the shear-driven, not the pressure-driven, scaling and
depart from Bernoulli by an amount set by the tangential impulse; the extent of gas action grows as
$\delta$ increases; and the translational anisotropy $\Tpar/\Tperp$ and its $\De_\theta=1$ surface are
measurable by molecular-beam time-of-flight, coinciding with $(z/D)_f(\delta)$. A Newtonian-gas CFD
\cite{Zahoor2018,Rubio2021,HerradaMontanero2016}, post-processed for $\De_\theta$ over its own field, will
show its shear zone overlapping $\De_\theta>0.05$ wherever $\delta$ is small---inconsistent with its own
solution; direct-simulation Monte Carlo \cite{Bird1994} is the arbiter in the transitional band.

We have treated the liquid as Newtonian. Sheathing the flow-focused jet in a dilute polymer
(polyethylene oxide) to stabilise and lengthen it has only just been introduced for the most demanding
SFX \cite{Vakili2026}, and it makes the liquid non-Newtonian as well: the gas keeps its deviatoric stress
directed while $\De_\theta\gtrsim1$, and within the high-shear surface region so created the polymer chains
stay stretched while $\mathrm{Wi}\gtrsim1$ (see \cite{Vakili2026}); the two under-relaxations stack. The kinetically computed
$\tau_s$ reported here is the input that closes that coupled problem, of which the present shear-driven,
kinetically resolved single-fluid result is the prerequisite.

\begin{acknowledgments}
This work was partially supported by the Spanish Ministry of Science, Innovation and Universities (grant no. PID2022-140951OB/AEI/10.13039/501100011033/FEDER, UE).

The author thanks the sample-delivery community whose vacuum-jet measurements motivated this analysis.
\end{acknowledgments}

\end{document}